\documentclass[preprints,article,accept,pdftex,moreauthors]{Definitions/mdpi} 
\usepackage{changes}
\firstpage{1} 
\pubvolume{1}
\issuenum{1}
\articlenumber{0}
\pubyear{2022}
\copyrightyear{2022}
\datereceived{} 
\dateaccepted{} 
\datepublished{} 
\hreflink{https://doi.org/} 
\pdfoutput=1

\Title{Thermodynamics of Hot Neutron Stars and Universal Relations}
\TitleCitation{Thermodynamics of Hot Neutron Stars and Universal Relations}

\Author{Pavlos Laskos-Patkos $^{1,\dagger,\ddagger,*}$, Polychronis S. Koliogiannis $^{1,\dagger,\ddagger,*}$\orcidB{}, Alkiviadis Kanakis-Pegios $^{1,\dagger,\ddagger}$\orcidC{}, and  Charalampos C. Moustakidis $^{1,\dagger,\ddagger}$\orcidD{}}

\AuthorNames{Pavlos Laskos-Patkos, Polychronis Koliogiannis, Alkiviadis Kanakis-Pegios and Charalampos Moustakidis}

\AuthorCitation{Laskos-Patkos, P.; Koliogiannis, P.; Kanakis-Pegios, A.; Moustakidis Ch.}

\address{%
$^{1}$ \quad Department of Theoretical Physics, Aristotle University of Thessaloniki, 54124 Thessaloniki, Greece; alkanaki@auth.gr (A.K.P.); moustaki@auth.gr (C.C.M.)}

\corres{Correspondence: plaskos@physics.auth.gr (P.L.P); pkoliogi@physics.auth.gr (P.S.K.)}

\firstnote{These authors contributed equally to this work.}

\abstract{Over the last few years, the detection of gravitational waves from binary neutron star systems has rekindled our hopes for a deeper understanding of the unknown nature of ultradense matter. In particular, gravitational wave constraints on the tidal deformability of a neutron star can be translated into constraints on several neutron star properties using a set of universal relations. Apart from binary neutron star mergers, supernova explosions are also important candidates for the detection of multimessenger signals. Such observations may allow us to impose significant constraints on the binding energy of neutron stars. The purpose of the present study is twofold. Firstly, we investigate the agreement of finite temperature equations of state with established universal relations. Secondly, we examine the possible existence of a universal relation between the binding energy and the dimensionless tidal deformability, which are the bulk properties connected to the most promising sources for multimessenger signals. We find that hot equations of state are not always compatible with accepted universal relations. Therefore, the use of such expressions for probing general relativity or imposing constraints on the structure of neutron stars would be inconclusive (when thermal effects are present). Additionally, we show that the binding energy and the dimensionless tidal deformability exhibit a universal trend at least for moderate neutron star masses. The latter allows us to  set bounds on the binding energy of a 1.4 $M_\odot$ neutron star using data from the GW170817 event. Finally, we provide a relation between the compactness, the binding energy and the dimensionless tidal deformability of a neutron star that is accurate for cold and hot isentropic equations of state.}


\begin{document}


\section{Introduction} \label{sec1}
 
Neutron stars are associated with some of the most violent phenomena in our universe~\cite{Shapiro-1983,Glendenning-2000,Haensel-2007,Zeldovich-71,Weinberg-72,Schutz-85,Bielich-2020}. Their birth, via a supernova (SN) explosion, leads to the emission of gigantic amounts of energy in the form of neutrinos \cite{Lattimer-1989}. The energy carried away by neutrinos is strongly correlated to a bulk neutron star property, the binding energy. This quantity is very sensitive to the equation of state (EOS) and therefore, remains uncertain. Additionally, the death of a neutron star in the case of merger event is also a fruitful source of information concerning the nuclear EOS. The emission of gravitational waves (GW) during the inspiral phase of a binary neutron star (BNS) merger yields important data connected to another bulk property, the dimensionless tidal deformability $\Lambda$. The fact that both of these quantities are dependent on the structure of the compact star raises questions about their possible correlation. If this is the case, it would correspond to a ``bridge'' between the formation and the destruction of a neutron star. Considering that an SN explosion and a merger are two of the most energetic events in our cosmos and promising sources for multimessenger signals, we may finally be able to unravel the underlying nature of dense nuclear matter.

The new era of multimessenger astronomy demands the accurate determination of the nuclear EOS at finite temperature. Both BNS mergers and SN explosions are important candidates for multimessenger detection. In the case of an SN event, the temperature of nuclear matter may reach values higher than 30 MeV (in the present study the temperature will be given in units of MeV). On the other hand, the exact temperature of neutron stars during the inspiral phase of a merger is still an open problem. There are several studies suggesting an important temperature increment by proposing a variety of different mechanisms  \cite{Meszaros-1992,Bildsten-1992,Aras-2019,Kochanek-1992,Lai-1994,Reisenegger-1994,Ho-1994,Lai-2006,Lai-2017}. The predicted temperature range is wide, starting from almost negligible heating and reaching values $T\gg$ 1 MeV \cite{Meszaros-1992,Bildsten-1992,Aras-2019,Kochanek-1992}. 

There are two main methods in order to incorporate temperature in calculations concerning neutron stars; the isothermal approach, where temperature is a constant throughout the star, and the isentropic (adiabatic) approximation where the entropy per nucleon is the constant parameter. However, which description is more realistic? The classical notion, naively, associates the concept of thermal equilibrium with isothermality. However, this is not the case in the framework of general relativity (GR). As Tolman and Ehrenfest \cite{Tolman-1930-a,Tolman-1930-b,Tolman-1930-c,Tolman-1934} showed for the scenario of an ideal gas, temperature is subjected to gravity (TE effect) and therefore, isothermality and thermal equilibrium are found in contradiction. After that realisation, a lot of studies were devoted to investigating this link between thermodynamics and GR \cite{Sorkin-1981,Gao-2011,Gao-2012,Lima-2019,Roupas-2013,Roupas-2015,Buchdahl-1949,Oppenheim-2003,Santiago-2018,Santiago-2019}. Sorkin et al. \cite{Sorkin-1981} managed to derive (for a perfect fluid) the relativistic hydrostatic equilibrium equation by maximising the total entropy of the configuration. Later on, Gao \cite{Gao-2011,Gao-2012} extended the work of Sorkin et al. \cite{Sorkin-1981} for the case of charged fluids. Lima et al. \cite{Lima-2019} in a recent study showed that the TE effect was actually valid only in the occurrence of perfect fluids satisfying the relation $\mu/T=$ constant, where $\mu$ and $T$ are the chemical potential and temperature, respectively. In any case, thermodynamics and GR are tightly connected. Since the phenomena that we aim to study are considered in a relativistic framework, temperature should be inserted in a way that satisfies the demands of GR.

In the past decades there has been an extensive study of universal relations concerning neutron stars \cite{Yagi-2013,Maselli-2013,Yagi-2015,Yagi-2017,Alexander-2019}. A universal relation, between two properties of a star, is one that holds for any given EOS. Such expressions are of most importance, since they allow us to gain information about the properties of neutron stars (such as the moment of inertia) by measuring the correlated quantities (such as the tidal deformability \cite{Yagi-2013}). Furthermore, these relations provide the opportunity for probing GR in a way that is independent of the employed EOS. From another point of view, the significance of universal relations can be understood if we consider that such expressions have been used in order to impose constraints on the EOS \cite{Abbott-2018,Tan-2022} (see for example Figure 2 of \cite{Tan-2022}). Specifically, universal relations can be used in order to convert GW observations (tidal deformability measurements) into mass--radius contours \cite{Tan-2022}. As we mentioned, thermal effects are present during the last orbits of a coalescing BNS system and the temperature predictions cover a wide interval, larger than three orders of magnitude. Therefore, if we want to impose robust constraints on the EOS, it is very important to establish whether finite temperature EOSs are in agreement with universal relations. Related studies that have applied temperature dependent equations of state to rotating and nonrotating proto-neutron and neutron star models, searching for possible universal relations and investigating their origin, suggest that universality is about the independence of EOS and not of thermodynamic conditions~\cite{Martinon-2014,Marques-2017,Raduta-2020,Khadkikar-2021}.

In a recent study, Reed and Horrowitz \cite{Reed-2020} discovered a linear correlation between the binding energy of a specific mass configuration and the dimensionless tidal deformability of a 1.4 $M_\odot$ neutron star, $\Lambda_{1.4}$. Therefore, the effective constraining of the dimensionless tidal deformation would immediately result in the constraining of the binding energy of a compact star. In this way, the observation of an SN and the accurate determination of the binding energy would provide important information about the interior of neutron stars. On the other hand, the knowledge of the dimensionless tidal deformability may shed light on the mechanics of an SN explosion. It is possible that this peculiar connection between the two quantities derives from a universal relation. The fact that GW detectors are rapidly developing and that BNS mergers and SN explosions are very promising candidates for multimessenger signals motivates us to explore a more general relation (universal rule) that connects the tidal deformability and the binding energy of a compact star.

Our motivation for this study is twofold: Firstly, we wish to investigate the compatibility of established universal relations with hot EOSs. If hot EOSs do not reproduce universal relations, then the use of such expressions for probing GR or constraining the nuclear EOS would be erroneous (when thermal effects are present). Secondly, we seek a possible universal rule that connects the binding energy to the dimensionless tidal deformability of a neutron star, which are the bulk properties associated with the two most relevant sources for multimessenger detections, SN explosions and BNS mergers.

This paper is organised as it follows.  Sections \ref{sec2}--\ref{sec5} are dedicated to the discussion of preliminary notions. Specifically, in Section \ref{sec2} we present the concept of the thermal equilibrium in GR, while in Sections \ref{sec3} and \ref{sec4} we discuss the binding energy and the neutron star tidal deformability, respectively. Additionally, in Section \ref{sec5} we focus on the description of an established universal relation (rescaled entropy vs. compactness \cite{Alexander-2019}), which involves the total entropy of a neutron star. Section \ref{sec6} is devoted to the presentation of our results and their implications. Finally, Section \ref{sec7} contains the concluding remarks of the present study.

\section{Thermal Equilibrium in General Relativity} \label{sec2}

The metric for a static and spherically symmetric star is 
\begin{equation} \label{eq1}
    ds^2 = g_{\mu\nu}dx^\mu dx^\nu = -e^{2\Phi(r)}dt^2+e^{2\lambda(r)}dr^2+r^2(d\theta^2+\sin^2\theta d\phi^2),
\end{equation}
where $\Phi,\lambda$ denote the metric functions that depend on the radial coordinate $r$. For such systems the condition for thermal equilibrium is given by the TE theorem \cite{Tolman-1930-a,Tolman-1930-b,Tolman-1930-c}
\begin{equation} \label{eq2}
    T(r) \sqrt{-g_{tt}(r)} = T(r) e^{\Phi(r)} = constant=T_0.
\end{equation}

This statement implies that temperature is subjected to gravity. Additionally, there is another similar expression that involves the chemical potential of a star, known as Klein's theorem \cite{Klein-1949}, and it states that $\mu(r)e^{\Phi(r)} =$ constant. The combination of these two theorems leads to a very elegant relation between the chemical potential and the temperature of the system which is 
\begin{equation} \label{eq3}
    \frac{\mu}{T} = constant.
\end{equation}

As it was shown by Lima et al. \cite{Lima-2019}, Equation (\ref{eq2}) of the TE theorem is actually valid only under certain assumptions. In particular, the fluid under consideration must be ideal and Equation (\ref{eq3}) has to be valid. The authors combined the energy conservation equation, the Gibbs law and the Euler relation, for a relativistic fluid, and studied the cases: (a) $\mu=$ 0 and (b) $\mu \neq $ 0 \cite{Lima-2019}. The former scenario led to a concrete proof of the TE theorem. In the latter case, they managed to establish a more general relation between the chemical potential and the temperature of the fluid
\begin{equation} \label{eq4}
    \frac{\partial \ln{(T\sqrt{-g_{tt}}})}{\partial x_i}+\frac{\mu}{T S_b}\frac{\partial \ln{(\mu\sqrt{-g_{tt}}})}{\partial x_i}=0,
\end{equation}
where $x_i$ corresponds to a spatial coordinate and $S_b$ is the entropy per nucleon \cite{Lima-2019}. It is worth pointing out that Equation (\ref{eq4}) can be thought of as a generalisation of the TE theorem that governs the thermal equilibrium of a self-gravitating object in GR.

\section{Binding Energy} \label{sec3}

The binding energy stands for the energy profit due to the assembling of \emph{N} baryons to form a stable star. As in the case of an atomic core, the binding energy can be evaluated via~\cite{Shapiro-1983,Glendenning-2000,Haensel-2007} 
\begin{equation} \label{eq5}
    E_b=M_bc^2-Mc^2=N m_b c^2-Mc^2,
\end{equation}
where $m_b$ is the average mass of a single nucleon, $M_b$ is the baryonic mass of the star and $M$ stands for the gravitational mass, which is given by the solution of Tolman--Oppenheimer--Volkoff (TOV) equations. Specifically, 
\begin{equation} \label{eq6}
    M = 4\pi \int_0^R r^2\rho (r)dr, 
\end{equation}
where $\rho(r)$ is the density profile and $R$ is the radius of the star. The total number of nucleons $N$ is found from the following integral
\begin{equation} \label{eq7}
    N=\int n(r) dV,
\end{equation}
where 
\begin{equation} \label{eq8}
    dV =4 \pi r^2 \sqrt{g_{rr}(r)} dr=4 \pi r^2 e^{\lambda(r)} dr,
\end{equation}
is the infinitesimal proper volume and 
\begin{equation} \label{eq9}
    e^{\lambda(r)}=\sqrt{g_{rr}(r)}=\left(1-\frac{2Gm(r)}{rc^2}\right)^{-1/2}.
\end{equation}

In Equation (\ref{eq9}), $m(r)$ stands for the mass distribution of the star.

There are several suggestions about which value of $m_b$ is more appropriate. Many authors assume that it corresponds to $\sim$ 939 MeV/$c^2$, which is the mass of protons and neutrons (approximately equal masses). Following the authors of Refs. \cite{Lattimer-2001,Reed-2020}, the average baryon mass is taken to be the mass of $^{56}$Fe/$56=930.412$ MeV/$c^2$.

\section{Tidal Deformability} \label{sec4}

Gravitational waves emitted during the last orbits of an inspiraling BNS system may offer important information, concerning the structure of neutron stars, through the measurement of a quantity known as tidal deformability  \cite{Flanagan-2008,Hindeder-2008,Damour-2009,Postnikov-2010,Hindeder-2010,Fattoyev-2013,Lackey-2015}.
Tidal deformability $\lambda$ corresponds to the proportionality coefficient between the induced quadrapole moment $Q_{ij}$ of the star and the external tidal field $\mathcal{E}_{ij}$ of its companion in a BNS system
\begin{equation} \label{eq10}
    Q_{ij} = -\lambda \mathcal{E}_{ij}.
\end{equation}

The evaluation of tidal deformability requires the knowledge of the second tidal Love number $k_2$, which is given by
\begin{equation} \label{eq11}
\begin{split}
    k_2 & = 
\frac{8C^5}{5}(1-2C^2)[2+2C(y_R-1)-y_R]
\times\{2C(6-3y_R+3C(5y_R-8))
\\
& +
4C^3(13-11y_R+C(3y_R-2)+2C^2(1+y_R))
\\
& +
3(1-2C)^2(2-y_R+2C(y_R-1)\log(1-2C))\}^{-1},
\end{split}
\end{equation}
where $C=GM/(Rc^2)$ is the star's compactness and $y_R=y(R)$ is extracted from the solution of the nonlinear differential equation
\begin{equation} \label{eq12}
    ry'(r)+y^2(r)+y(r)F(r)+r^2Q(r)=0,
\end{equation}
where
\begin{equation} \label{eq13}
    F(r)=\left[1-\frac{4\pi r^2G}{c^4}(\epsilon(r)-p(r))\right]\left(1-\frac{2Gm(r)}{rc^2}\right)^{-1},
\end{equation}
and 
\begin{equation} \label{eq14}
\begin{split}
    r^2Q(r) & = 
    \frac{4\pi r^2G}{c^4}\left[5\epsilon(r)+9p(r)+\frac{\epsilon(r)+p(r)}{c_s^2}\right]\times \left(1-\frac{2Gm(r)}{rc^2}\right)^{-1}
    \\
    &-
    6\left(1-\frac{2Gm(r)}{rc^2}\right)^{-1}-
    \frac{4m^2(r)G^2}{r^2c^4}\left(1+\frac{4\pi r^3p(r)}{m(r)c^2}\right)\left(1-\frac{2Gm(r)}{rc^2}\right)^{-2}.
\end{split}
\end{equation}

In Equations (\ref{eq13}) and (\ref{eq14}), $\epsilon$, $p$ and $c_s=$ $(\partial p $/$\partial \epsilon)^{1/2}$  stand for the energy density, pressure and speed of sound, respectively. Equation (\ref{eq12}) has to be integrated simultaneously with the TOV equations (for self-consistency)  and its boundary condition is $y(0)=2$.

The study of universal relations, which is the key ingredient of the present work, often involves dimensionless quantities. Thus, we define the dimensionless tidal deformability~as
\begin{equation}\label{eq15}
    \Lambda=\frac{2}{3}k_2C^{-5}.
\end{equation}

The dimensionless tidal deformability follows a series of unexpected universal relations, known as I-Love-Q and Love-C relations \cite{Yagi-2013,Yagi-2015,Yagi-2017}. The mentioned relations have several important applications as they allow us to impose constraints on the neutron star structure and the EOS \cite{Abbott-2018,Tan-2022}. Concerning other interesting applications of these universal expressions, a recent study employed the I–Love–Q \cite{Yagi-2013} and Love-C \cite{Maselli-2013} relations in order to impose constraints on extraspatial dimensions \cite{Chakravarti-2020}.

\section{Rescaled Entropy} \label{sec5}

As previously mentioned, our primary goal was to establish whether finite temperature EOSs satisfy universal relations. Such relations are of most importance as they reveal similarities among nuclear models, and they allow us to probe GR in a way that is independent of the EOS. One of the EOS-independent relations that we employ was discovered by Alexander et al. \cite{Alexander-2019}. The authors studied the possibility of a universal relation which involves the total entropy of a neutron star. Forming a dimensionless quantity, the rescaled entropy, they showed that it follows a universal relation with respect to the compactness for isotropic and anisotropic stars. The universality of this relation (but in the case of anisotropic stars) appeared to be increasing for neutron stars on the verge of collapse to a black hole. Additionally, as the compactness of the star approaches the value $C=1/2$, the rescaled entropy also tends (approximately) to that of a black hole \cite{Alexander-2019}.

Starting from preliminary notions, the first law of thermodynamics \cite{Alexander-2019} 
\begin{equation} \label{eq16}
    dE = TdS-pdV+\mu dN,
\end{equation}
solved for the total entropy for a spherically symmetric self-gravitating object, yields
\begin{equation} \label{eq17}
    S = \frac{1}{T_0} \int_0^R \left(-g_{tt}\right)^{1/2} (\epsilon+p-\mu n) dV.
\end{equation}

In Equation (\ref{eq17}), $T_0=T(r)e^{\Phi(r)}$ is the redshifted temperature and $dV$ is the infinitesimal proper volume. Under the assumption that Equation (\ref{eq2}) is valid, the redshifted temperature  is a constant and therefore, it is excluded from the integral.

It is proven that for studying neutron stars with cold EOS, one can employ the following relation (for a detailed proof see the Appendix of Ref.\cite{Alexander-2019})
\begin{equation} \label{eq18}
  \mu(R) \frac{n}{\epsilon+p}  \left(\frac{g_{tt}(R)}{g_{tt}(r)}\right)^{1/2} = \mathcal{C}_1, 
\end{equation}
where $\mathcal{C}_1$ is a constant. Combining Equation (\ref{eq18}) and Klein's theorem, the form of the total entropy is found to be
\begin{equation} \label{eq19}
    S =  \frac{4 \pi}{T'} \int_0^R \left(-g_{tt}g_{rr}\right)^{1/2} (\epsilon+p) r^2dr,
\end{equation}
where $T'=T_0/(1-\mathcal{C}_1)$. The rescaled entropy corresponds to the product of the total entropy and $T'$ divided by the star's mass
\begin{equation} \label{eq20}
    \frac{T'S}{Mc^2} = \frac{4 \pi}{Mc^2} \int_0^R \left(-g_{tt}g_{rr}\right)^{1/2} (\epsilon+p) r^2dr.
\end{equation}

Alexander et al. \cite{Alexander-2019} showed that, apart from the star's compactness, the rescaled entropy followed a universal relation with respect to the star's moment of inertia. The latter means that it will probably also be connected to all the quantities of the I-Love-Q universal relations \cite{Yagi-2013,Yagi-2015,Yagi-2017}.

\section{Results and Discussion} \label{sec6}
In this section, we present the nuclear model that we employ for the construction of hot neutron star models. Additionally, we compare the following two universal relations to the results from hot EOSs: (a) rescaled entropy vs. compactness \cite{Alexander-2019} and (b) dimensionless tidal deformability vs. compactness \cite{Maselli-2013}. We wish to establish whether universal relations can be employed for the probing of the nuclear EOS and GR even when thermal effects are present. Finally, we are going to investigate the possible existence of a universal connection between the binding energy and the dimensionless tidal deformability of a star.

\subsection{Hot Equations of State} \label{sub1}

In order to reproduce a universal relation, we used a set of cold EOSs. Information about most of the employed EOSs in this study can be found in Refs. \cite{Koliogiannis-2020,Costantinou-2014}. For the KASV EOS, the reader is referred to Ref. \cite{Kurkela-2014}.  Furthermore, we used the Tolman VII solution \cite{Tolman-1939}, which corresponds to a realistic analytical solution of the TOV equations. For the construction of the hot EOSs, we adopted the momentum-dependent interaction (MDI) model \cite{Prakash-1997}. The parametrisation of the MDI model was based on the data provided by Akmal et. al. \cite{Akmal-1998}. MDI+APR1, has been previously used for the study of the properties of cold and hot nuclear matter \cite{Koliogiannis-2020,Koliogiannis-2021,Kanakis-2021,Moustakidis-2008}.
In particular, for isothermal neutron stars the core was described by the $T = 20$ MeV MDI+APR1 EOS. In the case of isentropic configurations, we employed the MDI+APR1 EOS with $S_b =$ 1 $k_B$ and $S_b =$ 2 $k_B$, and three different lepton fractions $Y_l=$ 0.2, 0.3 and 0.4. For the crust region we employed two models. For the cold configurations, we used the EOS of Baym, Pethick and Sutherland \cite{Baym-1971}, whereas for the low-density regime in the case of hot neutron stars, we employed the Lattimer and Swesty EOS \cite{Lattimer-1991}. Details for extending the nuclear EOS at finite temperature can be found in Refs. \cite{Moustakidis-2008,Costantinou-2014,Constantinou-2015a,Constantinou-2015b,Wei-2021,Oertel-2017,Schneider-2019,Carbone-2019,Chesler-2019,Lattimer-2016}.

\begin{figure}[H]
    \includegraphics[width=1\textwidth]{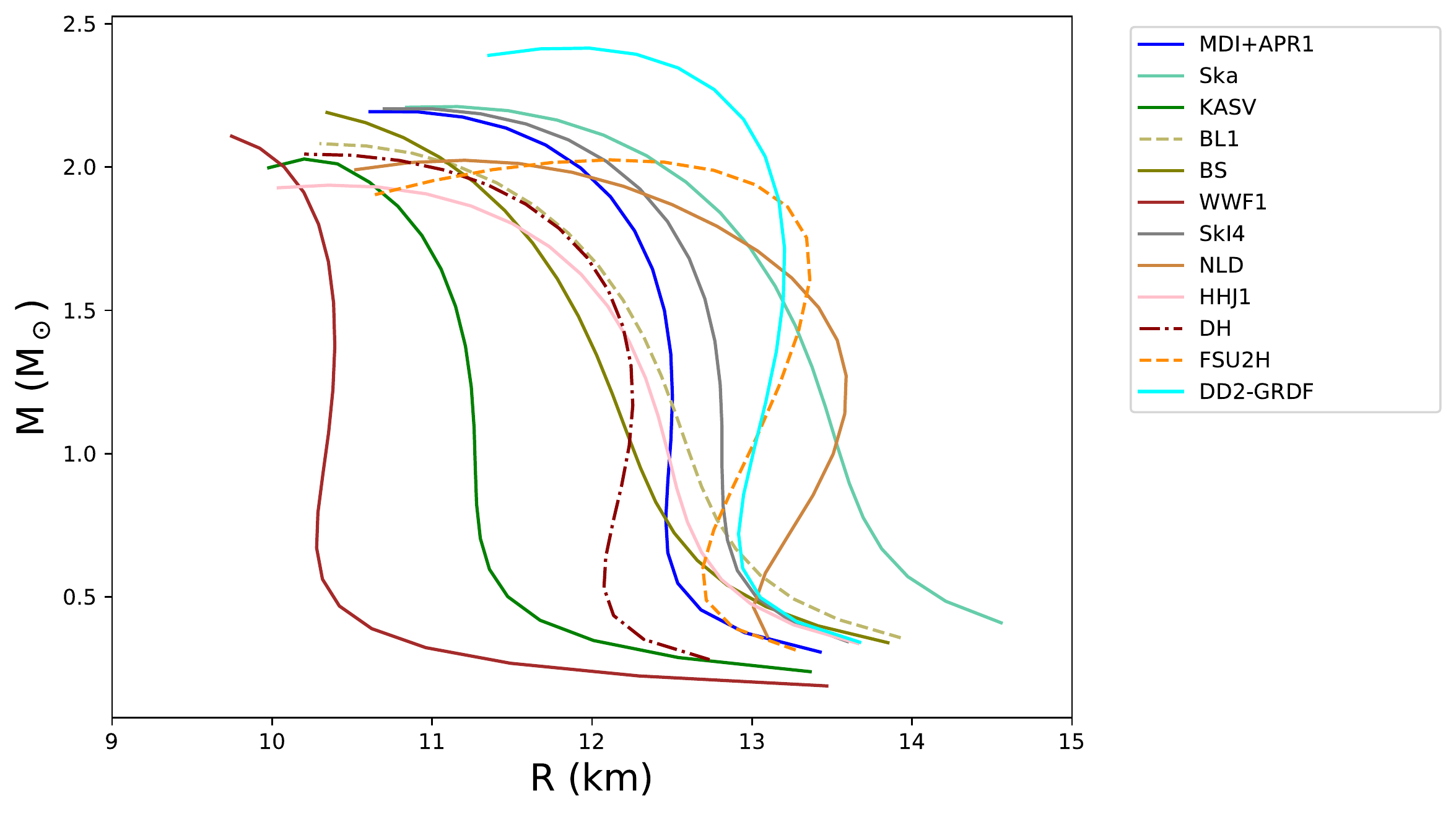}
    \caption{Mass--radius dependence for the employed cold EOSs.}
    \label{f1}
\end{figure}

\begin{figure}[H]
    \includegraphics[width=1\textwidth]{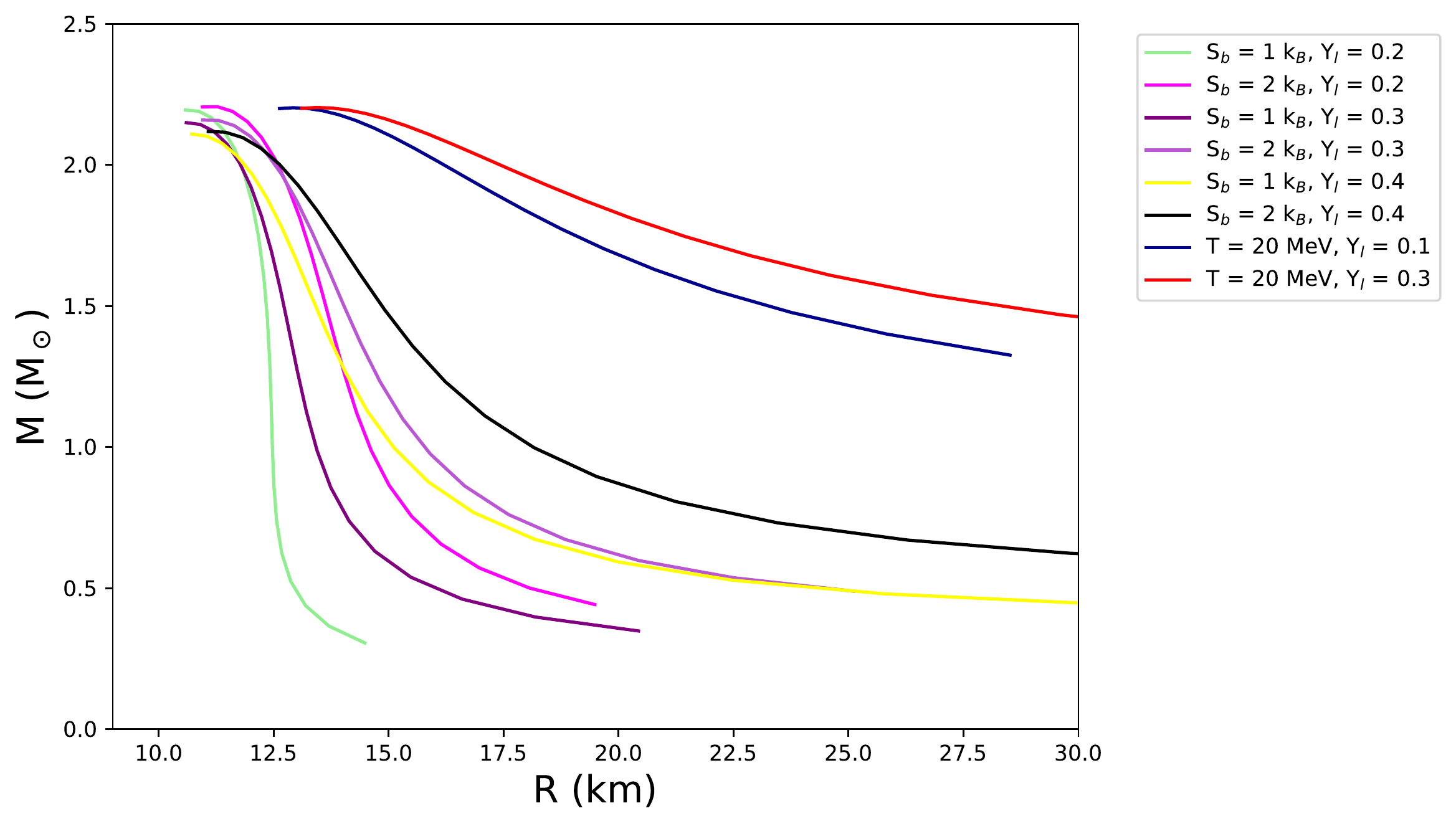}
    \caption{Mass--radius dependence for the MDI-APR1 EOS at finite temperature.}
    \label{f2}
\end{figure}

Figures \ref{f1} and \ref{f2} depict the mass--radius dependence for the cold and hot EOSs employed, respectively. Different EOSs produce a wide range of results for the M--R diagrams. As one can observe, in the case of a constant temperature, the maximum gravitational mass of the star, $M_{max}$, is not affected by the temperature. It is noteworthy that the main implications of a finite temperature are found in the size of the star. Specifically, the difference in radius of a cold neutron star and a hot one is dramatic. The radius of a 1.4 $M_\odot$ at $T=$ 20 MeV star is almost two times the radius of the corresponding cold configuration \cite{Koliogiannis-2021}. 

In the case of isentropic neutron stars, we find that for a specific lepton fraction, as the entropy per nucleon increases, the maximum gravitational mass also increases. It is notable, that even though the entropy doubles (from 1 $k_B$ to 2 $k_B$), the increment of $M_{max}$ is almost negligible. Once again, a finite temperature appears to have a strong impact on the radius of the star. Furthermore, we find that the lepton fraction also plays a crucial role on the bulk properties of the neutron star. Specifically, for configurations with the same entropy, as the lepton fraction increases, the maximum mass (slightly) decreases, whereas the size of the star increases \cite{Koliogiannis-2021}.

\subsection{Rescaled Entropy vs. Compactness} \label{sub2}
We investigated the agreement of hot EOSs with the universal relation between the rescaled entropy and the compactness \cite{Alexander-2019}. It is worth pointing out that in the case of hot neutron stars, the form of the  rescaled entropy is more complicated than the left-hand side (LHS) of Equation (\ref{eq20}). The reason is that, as Lima et al.~\cite{Lima-2019} showed, the TE effect holds only under very specific conditions (vanishing chemical potential or $\mu$/$T$ = constant). None of these conditions is satisfied for hot EOSs and the temperature is not sufficiently low to neglect its effects. In any case, for both cold and hot EOSs, we solved the TOV equations, and then we evaluated the integral in the right-hand side (RHS) of Equation (\ref{eq20}) (similar to the Ref. \cite{Alexander-2019}).

\begin{figure}[H] 
    \includegraphics[width=0.8\textwidth]{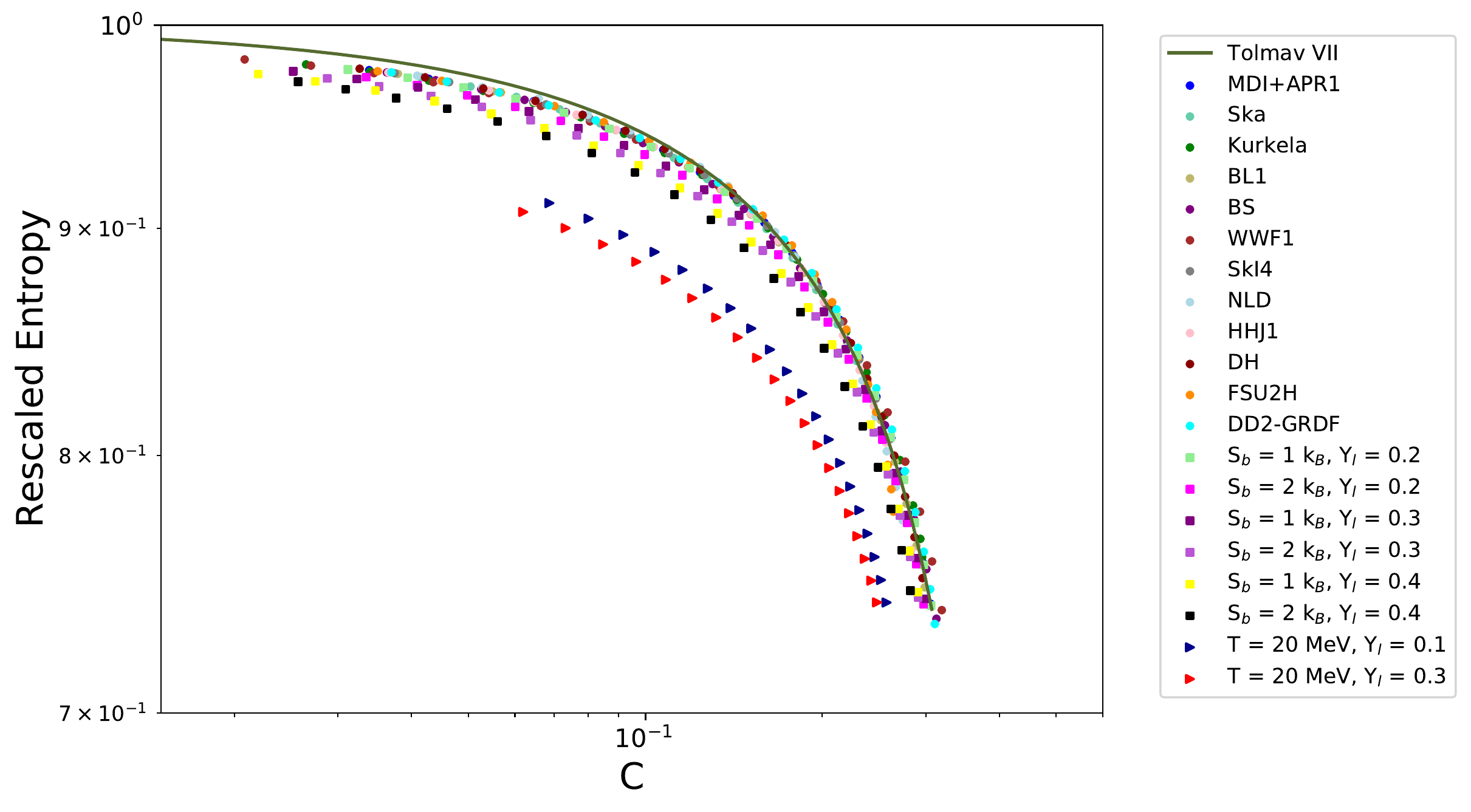}
    \caption{The rescaled entropy as a function of the star's compactness for cold and hot (isothermal and isentropic) EOSs. The circular points correspond to cold EOS, whereas the rectangular and the triangular points stand for isentropic and isothermal neutron stars, respectively. The solid line is the prediction of the Tolman VII solution.}
    \label{3}
\end{figure}

In Figure \ref{3}, we display the universal relation of Alexander et al. \cite{Alexander-2019} and its agreement with finite temperature EOSs. In comparison to the M--R diagrams, where the results for each EOS were presented using full lines, we now plot only a set of points (corresponding to different configurations) for practical purposes. As one can observe, the cold EOSs and the Tolman VII solution reproduce the universal relation found in Ref. \cite{Alexander-2019}. For low values of entropy per nucleon and lepton fraction, we found that isentropic EOSs also exhibited this universal behaviour. Actually, the lepton fraction appears to play a major role in the agreement with the universal trend. Note that even for the same value of entropy per nucleon, increasing the lepton fraction leads to larger differences from the prediction of cold EOSs. Even though isentropic EOSs gradually diverge from the universal relation, as the entropy and lepton fraction increase, they still form a narrow band. In contrast, isothermal EOSs diverge significantly from the curve. It needs to be mentioned that the temperature in the centre of the star, for the case of the isentropic EOSs, may reach very high values (up to 65 MeV), and therefore an argument of a low temperature hot EOS does not apply here.

The fact that isentropic EOSs are in relatively good agreement with the universal relation of Alexander et al. \cite{Alexander-2019} raises questions about the rescaled entropy. As previously mentioned, Lima et al. \cite{Lima-2019} predicted a more general form for the relation between $\mu$ and $T$, which is given by Equation (\ref{eq4}) and corresponds to the equation governing the thermal equilibrium of the system. For isentropic stars, we can show that, after some calculations,  
\\
\begin{equation} \label{eq21}
    \frac{\partial }{\partial x_i}\left[(TS_b+\mu)\sqrt{-g_{tt}}\right]=0,
\end{equation}
and therefore
\begin{equation} \label{eq22}
    (TS_b+\mu)\sqrt{-g_{tt}}={\cal C}_2, \hspace{0.5 cm} \frac{p+\epsilon}{n}\sqrt{-g_{tt}}={\cal C}_2.
\end{equation}

The value of the constant ${\cal C}_2$ can be evaluated from the boundary condition on the surface of the star. As in the case of cold EOSs (see Refs. \cite{Goldman-1989,Goldman-1990,Lattimer-2005}) we~obtain 
\begin{equation} \label{eq23}
{\cal C}_2 = m_b c^2 e^{\Phi(R)},
\end{equation}
where $m_b$ denotes the average mass of a nucleon.
Apparently, Equations (\ref{eq22}) stand for a general expression of the TE theorem, since if $\mu = $ 0 or $\mu / T =$ constant, Equation (\ref{eq22}) yields Equation (\ref{eq2}). 

Using the Euler relation \cite{Lima-2019}, the generalised rescaled entropy, which corresponds to the RHS of Equation (\ref{eq20}), is written as
\begin{equation}\label{eq24}
   \frac{1}{M c^2} \int (T S_b + \mu)\sqrt{-g_{tt}}ndV=\frac{(T S_b + \mu)}{M c^2}\sqrt{-g_{tt}} \int ndV=\frac{(T S_b + \mu)}{M c ^2}\sqrt{-g_{tt}}N.
\end{equation}

The result of Equation (\ref{eq24}) satisfies a universal relation with the star's compactness. Therefore, for stars with constant entropy per nucleon, Equation (\ref{eq24}) takes the form
\begin{equation} \label{eq25}
   \frac{(T S_b + \mu)}{Mc^2}\sqrt{-g_{tt}}\frac{S}{S_b}  = f(C),
\end{equation}
and then
\begin{equation} \label{eq26}
   \frac{Mc^2}{S}f(C)=\frac{(T S_b + \mu)}{S_b}\sqrt{-g_{tt}}=\frac{p+\epsilon}{nS_b}\sqrt{-g_{tt}}=\frac{m_bc^2\sqrt{1-2C}}{S_b},
\end{equation}
where $f(C)$ is independent of the employed EOS and $S=NS_b$ is the total entropy of the star.

It is easy to identify that the rescaled entropy is tightly connected to the ratio of baryonic and gravitational mass. From Equation (\ref{eq26}),
\begin{equation} \label{eq27}
   \frac{Mf(C)}{M_{b}}=\frac{(T S_b + \mu)
    }{m_b c^2}\sqrt{-g_{tt}}=\frac{p+\epsilon}{m_b n c^2}\sqrt{-g_{tt}}=e^{\Phi(R)}=\sqrt{1-2C}.
\end{equation} 

Then, we can express the universal relation for the rescaled entropy as a universal relation that involves the binding energy $(E_b)$ of a star
\begin{equation} \label{eq28}
   \frac{\sqrt{1-2C}M_{b}}{M}=\left(1+\frac{E_b}{Mc^2}\right)\sqrt{1-2C}=f(C).
\end{equation}

This is a very interesting result. Our study focused on a universal relation for the rescaled entropy, a quantity that involves the  thermodynamic variables of the system, and we concluded that the ratio of the baryonic and gravitational mass (multiplied by the redshift $(e^{\Phi(R)})$) should also satisfy the same relation with respect to compactness. From the fact that Equations (\ref{eq22}) hold not only for cold, but also for isentropic EOSs, we expect to
find similar results (to those for the rescaled entropy) if we evaluate the ratio of baryonic to gravitational mass multiplied by the redshift.

We need to comment that Lattimer and Prakash~\cite{Lattimer-2001} have already proposed a relation between the reduced binding energy $(E_{b}/Mc^{2})$ and the compactness parameter, according to the form

\begin{equation} \label{eq29}
    \frac{E_b}{Mc^2}=\frac{d_{1} C}{1-d_{2} C},
\end{equation}
where $d_{1}=0.6$ and $d_{2}=0.5$. Breu and Rezzolla~\cite{Breu-2016} revisited the coefficients of Equation~\eqref{eq29} as $d_{1}=0.6213$ and $d_{2}=0.1941$, using modern EOSs and the $2~M_{\odot}$ constraint. In addition, the latter authors also considered a relation of the form

\begin{equation} \label{eq30}
    \frac{E_{b}}{Mc^{2}}=d_{3}C + d_{4}C^{2},
\end{equation}
where a quadratic dependence on the compactness parameter is applied, and the coefficients are $d_{3}=0.619$ and $d_{4}=0.1359$. However, as the authors stated, Equation~\eqref{eq30} is marginally better than Equation~\eqref{eq29} and should be considered equivalent to Equation~\eqref{eq29}.

It has to be noted that Alexander et al.~\cite{Alexander-2019} and Lattimer and Prakash \cite{Lattimer-2001} discovered the same universal relation following an entirely different philosophy.

\subsection{Tidal Deformability vs. Compactness} \label{sub3}

\begin{figure}[H] 
    \includegraphics[width=1\textwidth]{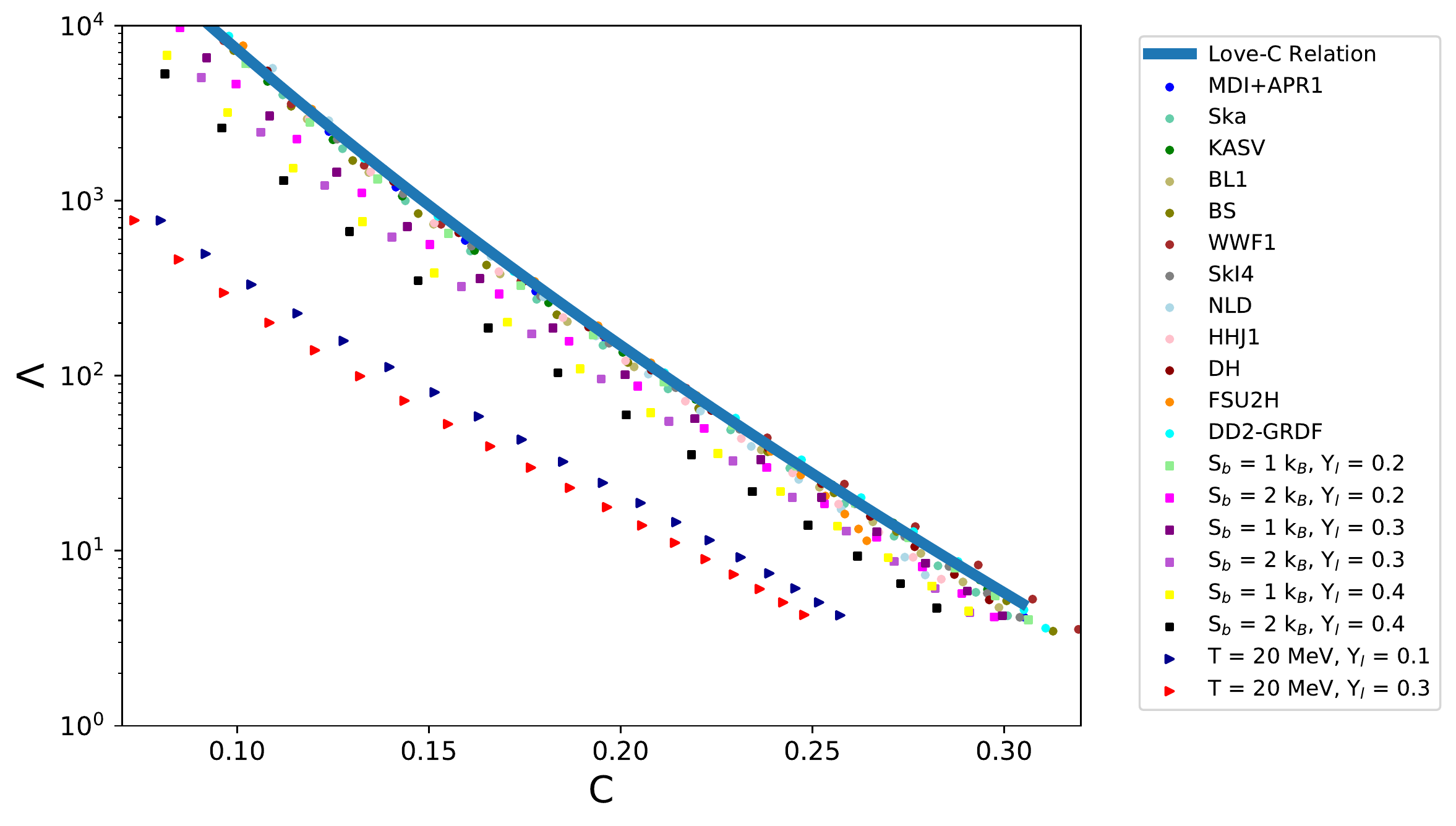}
    \caption{Universal relation connecting the tidal deformability and the star's compactness. The circular points correspond to cold EOS, whereas the rectangular and the triangular points stand for isentropic and isothermal neutron stars, respectively. The blue solid line corresponds to the Love-C fit of Maselli et al. \cite{Maselli-2013}. We employed the parameter set from the review study of Yagi and Yunes \cite{Yagi-2017}.}
    \label{f4}
\end{figure}

The importance of universal relations can be understood if we consider that such expressions are used in order to probe GR and to impose constraints on the nuclear EOS \cite{Abbott-2018,Tan-2022}. The second universal relation that we employed in order to test the behaviour of hot EOSs, is the one that connects the dimensionless tidal deformability of star to its compactness \cite{Maselli-2013,Yagi-2017}. In Figure \ref{f4}, we display this dependence for finite temperature EOSs as well. Apart from the predictions for a variety of cold and hot EOSs, we also plotted the fit found in Ref. \cite{Maselli-2013}. 

Once again, the universal relation is satisfied only in the case where the entropy per nucleon is 1 $k_B$ and the lepton fraction 0.2. All the other EOSs gradually diverge from the universal rule. It is noteworthy that when the entropy becomes 2 $k_B$ and the lepton fraction is 0.3 or 0.4, the dimensionless tidal deformability (for a specific compactness) is less than half of the prediction from the universal relation. We need to underline that a value of 0.4 for the lepton fraction is probably too large in the case of a BNS merger and it is more realistic in the case of a core-collapse SN. Additionally, the dimensionless tidal deformability of isothermal neutron stars ($T$ = 20 MeV) is found to diverge from the prediction of the Love-C relation by several orders of magnitude. 

To sum up, in the case of a neutron star with relatively low entropy ($S_b$ = 1 $k_B$) and proton fraction ($Y_l=$ 0.2), a hot EOS is also in good agreement with the $\Lambda$--$C$ universal relation. In that scenario, this specific universal relation could be used for probing the theory of gravity or for imposing constraints on nuclear matter. If either the temperature or the proton fraction is higher, the divergence from the universal trend is very significant. Furthermore, isothermal EOSs cannot reproduce the universal relations under consideration. In any case, since the temperature range during the inspiral phase of a merger is not well-constrained (see Ref. \cite{Kanakis-2022} and references therein), it is difficult to draw solid conclusions. However, our results indicate the need for narrowing down the temperature range and also clarifying the thermodynamic conditions of neutron stars during the inspiral phase of a merger.

\subsection{Binding Energy vs. Tidal Deformability} \label{sub4}

As previously mentioned, one of our primary goals was to establish a universal relation between the dimensionless tidal deformability and the binding energy of a neutron star. These two bulk properties, which are highly sensitive to the EOS, are connected with the two most promising sources for multimessenger observations, SNs and BNS mergers. Actually, the authors in Ref. \cite{Reed-2020} discovered a linear correlation between the dimensionless tidal deformability and the binding energy, but only for specific mass configurations. In particular, they used only the dimensionless tidal deformability of a 1.4 $M_\odot$ neutron star, which restricts us in a very narrow regime of the $\Lambda$ axis, and there is no actual evidence that this correlation would extend for other intervals of $\Lambda$. We estimate that the correlation found in Ref. \cite{Reed-2020} rises from the fact that $\Lambda$ and $E_b$ are universally connected. Specifically, if we consider that there is a universal relation for the dimensionless tidal deformability with respect to the star's compactness \cite{Yagi-2017}, and also an approximate universal relation for the binding energy divided by the gravitational mass with the compactness \cite{Lattimer-2001}, there should be a universal rule connecting these two bulk properties. 

\begin{figure}[H] 
    \includegraphics[width=1\textwidth]{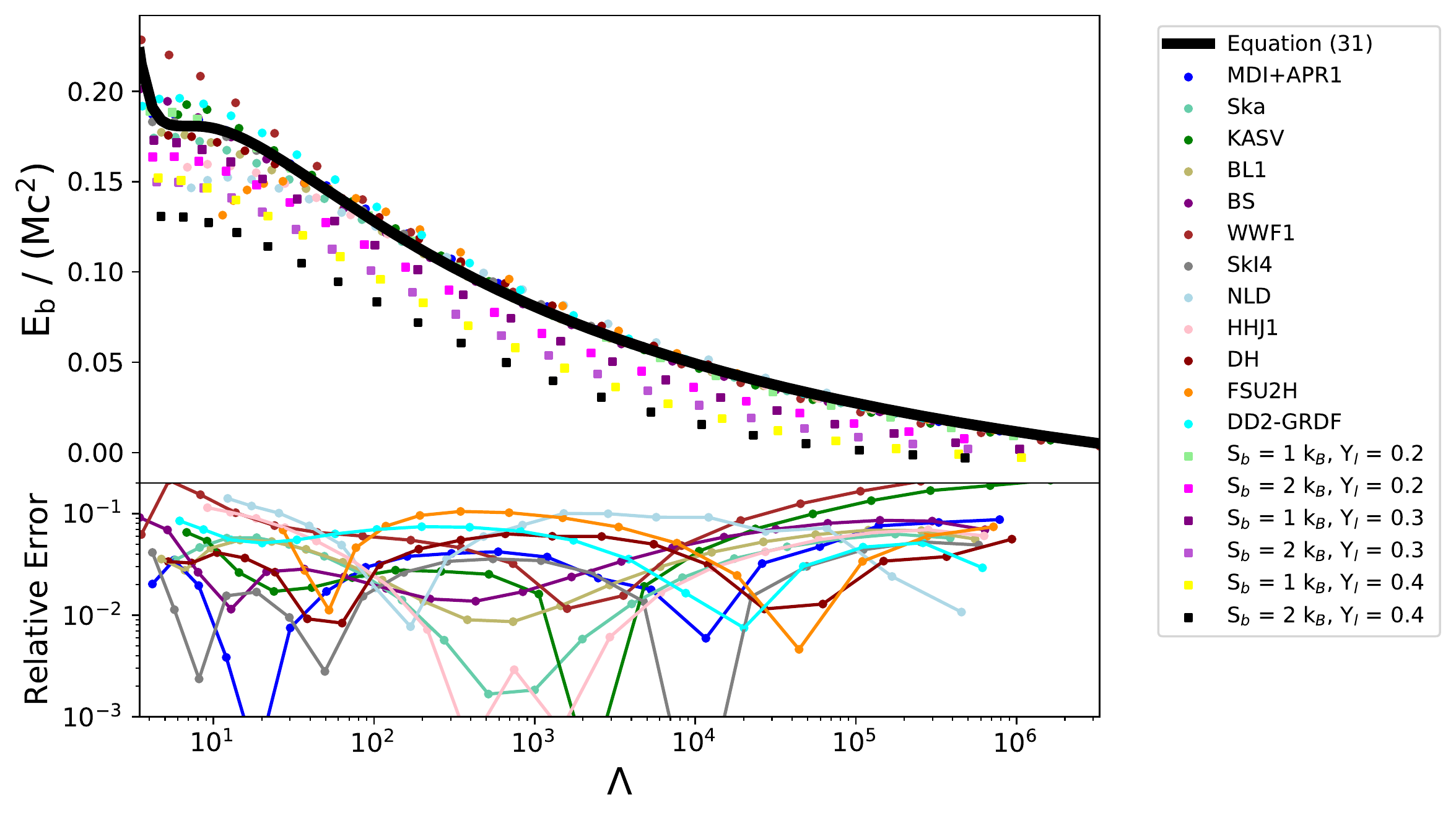}
    \caption{Top panel: The relation between the tidal deformability and the star's binding energy for cold and hot neutron stars. The circular points correspond to cold EOS, whereas the rectangle points correspond  to isentropic neutron stars. The solid line stands for the fit of Equation (\ref{eq31}), which is performed for cold EOSs. Bottom panel: Relative error for the fit of Equation (\ref{eq31}).}
    \label{f5}
\end{figure}

In Figure \ref{f5}, we display our results for the $E_b$/$(Mc^2)$ and $\Lambda$ dependence for cold and hot neutrons stars. First of all, it is important to note that the cold neutron stars in fact follow an approximate universal relation. For the trend of the cold EOSs, we provided a fit of the following form
\begin{equation} \label{eq31}
    \frac{E_b}{Mc^2} = \sum_{k=0}^3 \frac{a_k}{(\ln{\Lambda})^k}.
\end{equation}

The best fit values for the constants and the $ R^2$ index value (for goodness of fit) are presented in Table \ref{tab1}. Even though Figure \ref{f5} includes configurations beyond the maximum mass limit (for each EOS-unstable regime), for the evaluation of Equation (\ref{eq31}), only the stable stars were considered. 

\begin{table}[H] 
\caption{Fit parameters for the $E_b$/$(Mc^2)$--$\Lambda$ relation of Equation (\ref{eq31}). The $R^2$ index is also included for completeness. \label{tab1}}
\newcolumntype{C}{>{\centering\arraybackslash}X}
\begin{tabularx}{\textwidth}{CCCCC}
\toprule
\boldmath{$a_0$}	& \boldmath{$a_1$}	& \boldmath{$a_2$} 	& \boldmath{$a_3$} & \boldmath{$R^2$} 
\\
\midrule
$-$0.08399 & 1.52078 & $-$2.91006 & 1.85495 & 0.987 \\
\bottomrule
\end{tabularx}
\end{table}

Our results indicate that the linear dependence for specific configurations found in Ref. \cite{Reed-2020} is probably due to the fact that the binding energy and dimensionless tidal deformability are universally dependent. Additionally, the present fit is not limited to a specific  configuration (for each EOS)  and also extends
the range of configurations that had already been studied. It is notable that the universality starts to fail as a star reaches its stability limits. In that case, the margin of error for the prediction of the binding energy via the tidal deformation would increase. We also find that the predictions from hot EOSs gradually disagree with the universal fit as the entropy per nucleon and the lepton fraction increase. This raises an issue since an appropriate EOS for the simulations of core-collapse SNs should be characterised by a high temperature and high lepton fraction. As previously mentioned, the temperature range for a neutron star during the inspiral phase of merger covers three orders of magnitude. Once again, our results reveal the necessity of studying the thermal conditions in the inspiral phase of a BNS merger.

\begin{figure}[H] 
    \includegraphics[width=1\textwidth]{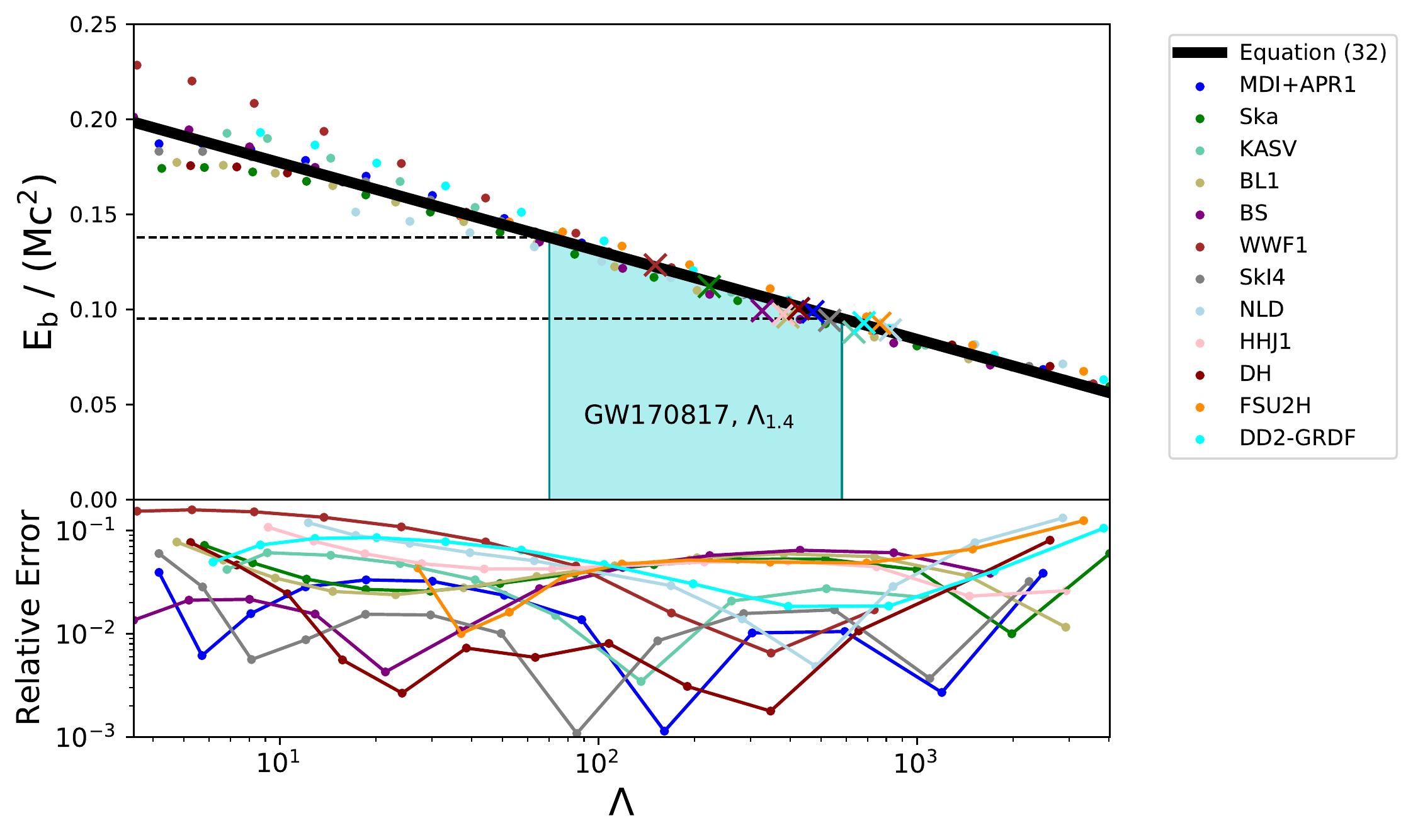}
    \caption{Top panel: Relation between the dimensionless tidal deformability and the star's binding energy for cold neutron stars and linear fitting. The circular points correspond to cold EOS. The solid line stands for the fit of Equation (\ref{eq32}), which was performed for cold EOSs, in the mass range $M\in\left[1M_\odot,M_{max}\right]$. The cross points represent the 1.4 $M_\odot$ configuration for each EOS. The shaded area shows the upper and lower values of the dimensionless tidal deformability ($\Lambda_{1.4}=190^{+390}_{-120}$) of a 1.4 $M_\odot$ neutron star from the analysis of the GW170817 event \cite{Abbott-2018}. Bottom panel: Relative error for the fit of Equation (\ref{eq29}).}
    \label{f6}
\end{figure}

The masses of the neutron stars that we observed lay in the range $M\in\left[1M_\odot,M_{max}\right]$. In particular, the lowest neutron star mass that was ever detected is equal to $M=1.174 \pm 0.004M_\odot$ \cite{Martinez-2015}. For that interval, we can fit a quite simple linear expression for the dependence between $E_b$/$(Mc^2)$ and $\Lambda$. In Figure \ref{f6}, we depicted the connection between the dimensionless tidal deformability and the binding energy for cold EOSs and for the mentioned mass range. We also provided a linear fit given by the following equation
\begin{equation}\label{eq32}
    \frac{E_b}{Mc^2}=b_0+b_1\ln \Lambda.
\end{equation}

\begin{table}[H] 
\caption{Fit parameters for the $E_b$/$(Mc^2)$--$\Lambda$ linear relation of Equation (\ref{eq32}). The $R^2$ index is also included for completeness. \label{tab2}}
\newcolumntype{C}{>{\centering\arraybackslash}X}
\begin{tabularx}{\textwidth}{CCC}
\toprule
\boldmath{$b_0$}	& \boldmath{$b_1$}	&  \boldmath{$R^2$} \\
\midrule
0.22350 & $-$0.02017 &  0.960 \\
\bottomrule
\end{tabularx}
\end{table}

The values of the constants $b_0$ and $b_1$ are included in Table \ref{tab2}. Once again, the universality starts to fail as configurations reach their maximum mass. In addition, Figure \ref{f6}  includes the constraints on the dimensionless tidal deformability ($\Lambda_{1.4}=190^{+390}_{-120}$) of a 1.4 $M_\odot$ neutron star, from the analysis of the GW170817 event (LIGO/Virgo collaboration) \cite{Abbott-2018}. The latter allows us the set a lower and an upper limit on the binding energy of a 1.4 $M_\odot$ compact star, using Equation (\ref{eq32}). The bounds on the binding energy can be found in Table \ref{tab3}. 

\begin{table}[H] 
\caption{Bounds on the binding energy of a 1.4 $M_\odot$ neutron star, using Equations (\ref{eq32}) and the data for $\Lambda_{1.4}$ from the analysis of the GW170817 event \cite{Abbott-2018}. The 1$\sigma$ error for the fit is also included.\label{tab3}}
\newcolumntype{C}{>{\centering\arraybackslash}X}
\begin{tabularx}{\textwidth}{CCC}
\toprule
\boldmath{$\Lambda_{1.4}$}  &\boldmath{${E_b}/{(Mc^2)}$}   
&\boldmath{$E_b$ $(10^{53}erg)$} \\
\midrule
70 & $0.1378\pm3.1\times10^{-3}$ & $3.4489\pm0.0793$  \\
580 &  $0.0952\pm3.9\times10^{-3}$ & $2.3814\pm0.0977$  \\
\bottomrule
\end{tabularx}
\end{table}

Constraining the binding energy gives us the opportunity to impose bounds on the energy emitted during an SN explosion, in the form of neutrinos. If the energy measured was less than the lower limit, it would be possible that a part of the binding energy had transferred into unobserved exotic particles \cite{Reed-2020}. Additionally, such an observation would be a chance to probe GR and test alternative theories of gravity. 
It is important to comment that the dimensionless tidal deformability has a direct dependence on the stiffness of the EOS through the compactness $C$ (see Equation (\ref{eq15})) and also an indirect dependence through the speed of sound (see Equation (\ref{eq14})). Therefore, in the case where the observed energy (in an SN explosion) overcomes the corresponding higher limit, it would be an indication of a softening of the EOS, possibly due to a phase transition.  In any case, it is worth pointing out that a disagreement with the predictions of Equation (\ref{eq32}) may result from the presence of finite temperature effects. As we have seen, hot EOSs do not obey the same universal relations as the cold ones.

\subsection{Binding Energy vs. Compactness}
\begin{figure}[H] 
    \includegraphics[width=0.95\textwidth]{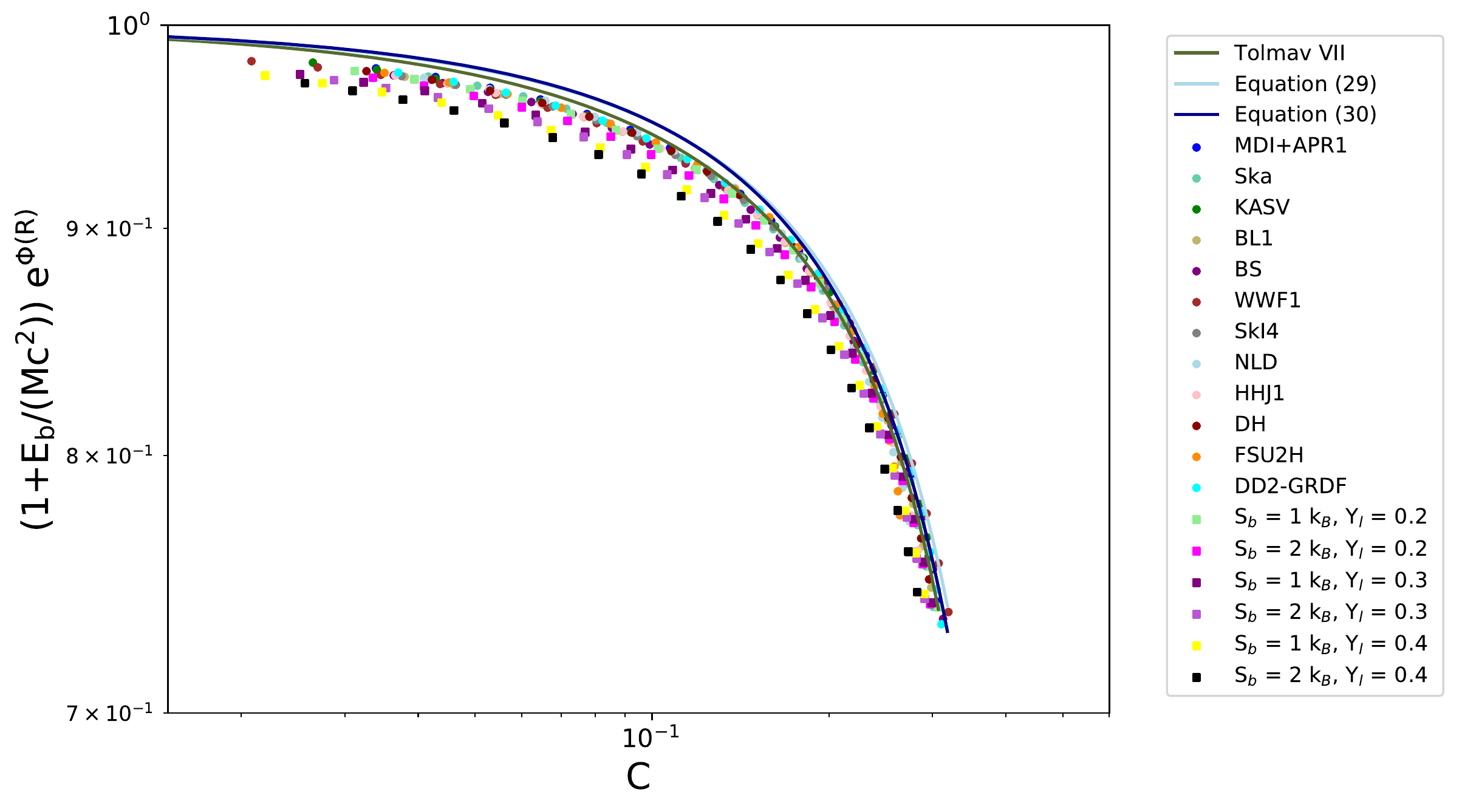}
    \caption{Universal relation between the neutron star compactness and the rescaled binding energy. The circular and the rectangular points stand for cold and isentropic EOSs, respectively. The corresponding fit of Lattimer and Prakash \cite{Lattimer-2001} of Equation (\ref{eq29}) and the fit of Breu and Rezzolla of Equation~(\ref{eq30}) are also plotted for completeness.}
    \label{f7}
\end{figure}

At this point, we wish to eliminate the differences of EOSs from the universal rule, which are present a) in the maximum mass regime for the cold cases and b) in the entire region for the hot cases. We showed that the rescaled entropy was equal to a quantity that involved the binding energy of the star (see Equation (\ref{eq28})). From now on, we refer to this quantity as the rescaled binding energy. Therefore, the rescaled binding energy should follow a universal relation with compactness as well. In Figure \ref{f7}, we present our results for the universal relation predicted from our empirical and theoretical analysis. The Lattimer and Prakash \cite{Lattimer-2001} fit of Equation (\ref{eq29}) and the corresponding quadratic approximation (Equation (\ref{eq30})) of Breu and Rezzolla \cite{Breu-2016} were also plotted for completeness. In fact, the universal relation appears to be the same as the one for the rescaled entropy and it is valid for cold EOSs as predicted. Additionally, isentropic EOSs are in relatively good agreement with the universal trend. The main conclusion is that the universal relation found in Ref. \cite{Lattimer-2001} coincides with the relation found in Ref. \cite{Alexander-2019}. The different approaches of these two studies, which produce the same results, raise some interesting questions. As shown by Ref. \cite{Alexander-2019}, the rescaled entropy exhibits an extraordinary behaviour at the transition from a neutron star to a black hole. We also proved that the rescaled entropy corresponded to a ratio of baryonic and gravitational mass multiplied by the star's redshift. Therefore, what is the role of the baryonic/gravitational mass ratio for a neutron star on the verge of collapse?

The reason that we performed the previous computation is not straightforward. In Ref. \cite{Alexander-2019}, the authors presented a universal relation between the neutron star's moment of inertia and the rescaled entropy. Thus, if we consider the I-Love-Q relations \cite{Yagi-2013}, there should be a connection between the  rescaled entropy and the tidal deformability. It is important to note that since Alexander et al. \cite{Alexander-2019} were aiming to study the transition from a neutron star to a black hole, they also included unstable configurations. The latter led to the conclusion that the universal relation between the dimensionless tidal deformability and the rescaled entropy (or rescaled binding energy) should also be accurate around the maximum mass limit. Additionally, hot EOSs are in relatively good agreement with the universal relation that connects the rescaled binding energy to the compactness of a star. In any case, it is interesting to examine the universal relation between the rescaled binding energy and the dimensionless tidal deformability.

\subsection{Binding Energy, Compactness and Tidal Deformability}

\begin{figure}[H] 
    \includegraphics[width=0.9\textwidth]{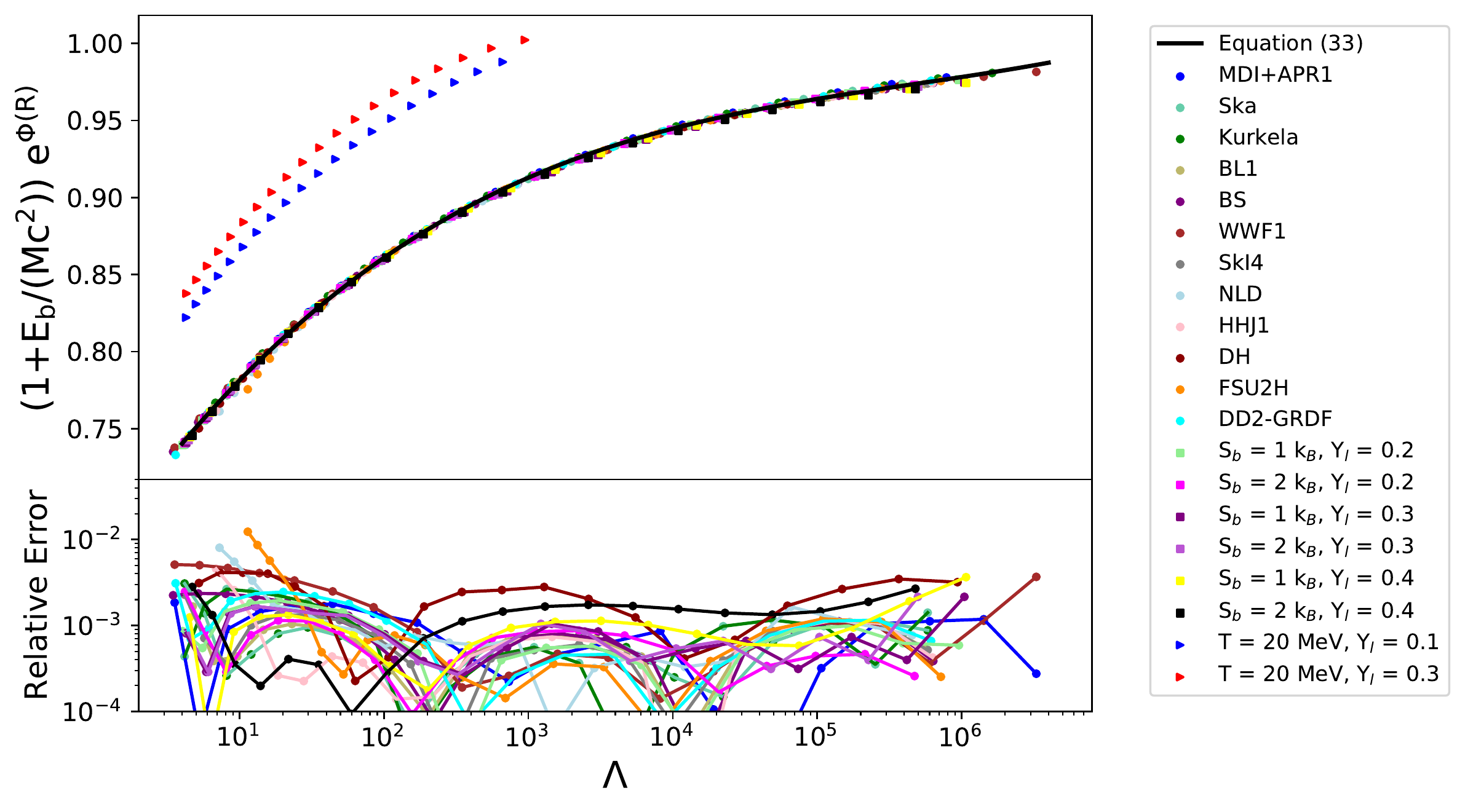}
    \caption{Top panel: Universal relation connecting the dimensionless tidal deformability, the star's binding energy and the compactness. The universality extends to finite isentropic EOSs as well. The circular and the rectangular points correspond to cold and isentropic EOSs, respectively. The triangular points correspond to isothermal EOSs. The solid line stands for the fit of Equation (\ref{eq33}), which was performed for cold and hot EOSs. Bottom panel: Relative error for the fit of Equation (\ref{eq33}).}
    \label{f8}
\end{figure}

In Figure \ref{f8}, we display another universal relation that correlates the binding energy of a star with its dimensionless tidal deformability. A fit of the form 
\begin{equation} \label{eq33}
    \left(1+\frac{E_b}{Mc^2}\right)\sqrt{1-2C}=\sum_{k=0}^3 C_k (\ln \Lambda)^k
\end{equation}
is also plotted for comparison. The constants, evaluated using the least squares method, are included in Table \ref{tab4}.

\begin{figure}
    \includegraphics[width=0.9\textwidth]{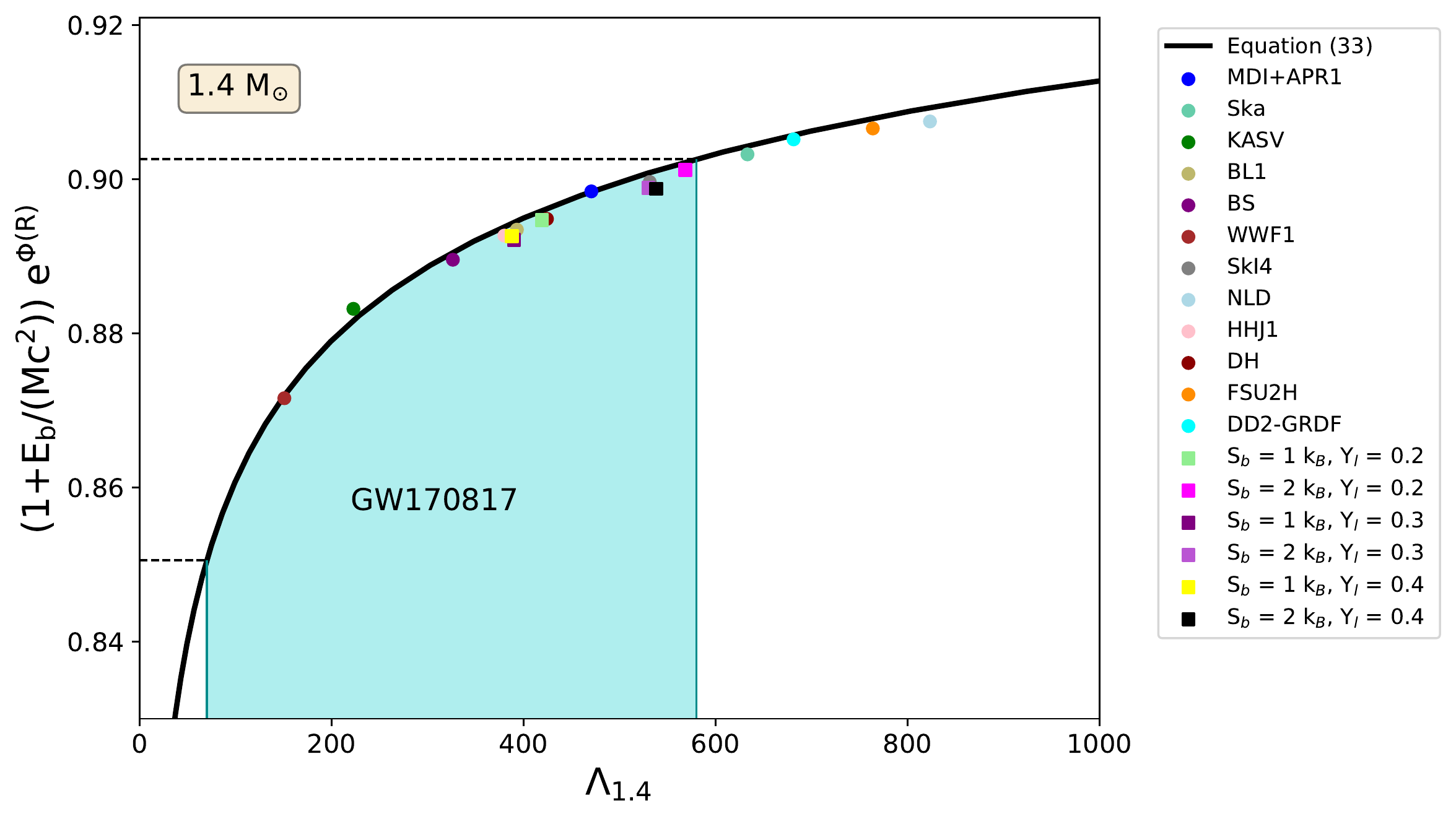}
    \caption{Universal relation connecting the dimensionless tidal deformability, the star's binding energy and the compactness. The plot contains only the data for the 1.4 $M_\odot$ configuration of each EOS. The circular and the rectangular points correspond to cold and isentropic EOSs, respectively. The solid line stands for the fit of Equation (\ref{eq33}), which is performed for cold and hot EOSs. The shaded area shows the upper and lower values of the dimensionless tidal deformability ($\Lambda_{1.4}=190^{+390}_{-120}$) of a 1.4 $M_\odot$ neutron star from analysis of GW170817 event \cite{Abbott-2018}. }
    \label{f9}
\end{figure}

We find that our approach does not only eliminate the spreading of predictions around the maximum mass configurations (including the unstable region), but also cancels the large differences that isentropic EOSs exhibit. We have managed to connect the compactness and the two bulk properties of a neutron star, which are connected to the most promising candidates for multimessenger signals, in a way that is independent of the cold EOSs and also holds in the scenario of isentropic neutron stars (in comparison to the case of $E_b$/$(Mc^2)$--$\Lambda$ and $\Lambda$--$C$ dependence). Isothermal EOSs are found in disagreement with the universal relation, and this is due to the fact that Equations (\ref{eq22}) and (\ref{eq23}) are not satisfied for constant temperature. Note that the precision of this specific universal relation is excellent, as the maximum relative error is found to be less than 2\% (considering only stable configurations the error is less than 0.6\%). In Figure \ref{f9}, we display the relation between the rescaled binding energy and the dimensionless tidal deformability but only for 1.4 $M_\odot$ neutron stars. The shaded region of the plot corresponds to the bounds on $\Lambda_{1.4}$ from the analysis of the GW170817 event \cite{Abbott-2018}. These bounds give us the opportunity the set lower and upper limits on the rescaled binding energy. Then, these limits translate into relations for the binding energy and the radius of a 1.4 $M_\odot$ neutron star. Hence, possible constrains on the radius of a 1.4 $M_\odot$ neutron star may allow us to gain information on the binding energy and vice versa.

\begin{table}[H] 
\caption{Fit parameters for Equation (\ref{eq33}). The $R^2$ index is also included for completeness.\label{tab4}}
\newcolumntype{C}{>{\centering\arraybackslash}X}
\begin{tabularx}{\textwidth}{CCCCC}
\toprule
\boldmath{$C_0$}	& \textbf{$C_1$}	& \boldmath{$C_2$} & \boldmath{$C_3$} & \boldmath{$R^2$} \\
\midrule
0.66656 & 0.05855 & $-$0.00402 & 0.00010 & 0.99969 \\
\bottomrule
\end{tabularx}
\end{table}
\vspace{-9pt}

From a lack of a solid theoretical argument on why the inclusion of the redshift factor yields the desired universality (which is strange) a series of interesting questions rise. As we saw, finite temperature affects the compactness of the compact star. Does this effect occur in a systematic way? Is this effect somehow cancelled when the redshift is taken into account? All of the above correspond to interesting future work suggestions. Additionally, we wish to extend this study in order to include other stable branches of compact stars, such as quark and twin stars. Finally, it would be interesting to investigate the form of the $E_b$--$\Lambda$ dependence in the framework of alternative theories of gravity.
\section{Conclusions} \label{sec7}
The present work was dedicated to the study of hot nuclear matter and EOS independent relations. In particular, we investigated the possibility of various universal relations, which are an intrinsic characteristic of GR. The main purpose was to establish whether hot EOSs satisfied the same universal relations as the cold ones or not. The importance of this study becomes clear if one considers the fact that there are several astrophysical phenomena (such as SNs or BNS mergers) where temperature may significantly affect the bulk properties of a neutron star. Since universal relations (concerning bulk neutron star properties) are used in order to probe GR and to impose constrains on the EOS,  we needed to establish if finite temperature models satisfied them as well. Furthermore, we searched for a possible universal rule between the binding energy of a neutron star and its tidal deformability. These two bulk properties are tightly connected with the most important candidates for multimessenger signals (SNs and BNS mergers).

Firstly, we confirmed the recent finding of Ref. \cite{Alexander-2019}, concerning the universal dependence between the total entropy of a neutron star and the compactness. It is notable that only the cold EOSs produced the mentioned universality. Isentropic EOSs were in good agreement as well, especially for a relatively low entropy and lepton fraction. In contrast, the isothermal ones did not respect the universality. It is noteworthy that if we considered the statement of the TE theorem (which predicts temperature gradient), isentropic EOSs were a more realistic approach for the description of hot neutron stars. Afterwards, using the generalized condition for thermal equilibrium (generalization of the TE effect) in GR \cite{Lima-2019} and the universal relation between the rescaled entropy and the compactness, we managed to show that the rescaled entropy was connected with the binding energy of a neutron star. Therefore, we found that a universality characterised the relation between the binding energy of a neutron star and its compactness. Of course, similar expressions have been proposed in previous studies \cite{Lattimer-2001}. In this way, we showed that the universal relations predicted in Refs. \cite{Alexander-2019,Lattimer-2001} were the same. Furthermore, we studied the EOS-independent relation between the tidal deformability and the compactness \cite{Yagi-2017}. We found that as the temperature and the lepton fraction rose, the differences between the fit and the predictions of the hot EOSs increased rapidly. Our results revealed the need for an accurate determination of the temperature range for a neutron star during the inspiral period of a merger.

Secondly, we investigated the possibility of a universal relation between the tidal deformability of a neutron star and its binding energy. We found that there was in fact a large region where these two quantities were connected. Our results indicated that the peculiar correlation found for the tidal deformability and binding energy of specific configurations in Ref. \cite{Reed-2020} was due to this universal connection. Furthermore, we pointed out that this universality started to break as the star approached its stability limits and it significantly failed for unstable configurations. Additionally, we found that as the temperature and the lepton fraction increased, the hot EOSs gradually differed from the predictions of the universal relation. It is noteworthy that we provided a linear fit for the dependence of $E_b$/$(Mc^2)$ and $\Lambda$ (for cold neutron stars), which allowed us to set bounds on the energy emitted during an SN explosion. Possible constraints on the binding energy from observations of tidal deformability may help to understand better the SNs explosion mechanics and the subsequent  neutrino emission. Furthermore, by exploiting the fact that the rescaled entropy follows a universal rule with the moment of inertia  \cite{Alexander-2019} (even for unstable configurations), its connection with the binding energy and also the I-Love-Q relations \cite{Yagi-2013,Yagi-2017}, we were able to predict another universal relation that associated the tidal deformability of a neutron star and its binding energy. We found that this universal expression was not only accurate in the unstable regime but also included hot isentropic EOSs. Therefore, provided there is a way to measure the rescaled binding energy, we found an appropriate universal relation for probing GR even when thermal effects are present. 

\vspace{6pt} 



\authorcontributions{Conceptualisation, P.L.-P., P.S.K., A.K.-P. and C.C.M.; methodology, P.L.-P., P.S.K. and A.K.-P.; software, P.L.-P., P.S.K. and A.K.-P.; validation, P.L.-P., P.S.K., A.K.-P. and C.C.M.; formal analysis, P.L.-P.,  P.S.K. and A.K.-P.; investigation, P.L.-P., P.S.K., A.K.-P., and C.C.M.; data curation, P.L.-P., P.S.K. and A.K.-P.; writing---original draft preparation, P.L.-P., P.S.K., A.K.-P., and C.C.M.; writing---review and editing, P.L.-P., P.S.K., A.K.-P. and C.C.M.; visualisation, P.L.-P., P.S.K., A.K.-P. and C.C.M.; supervision, C.C.M.; project administration, C.C.M.; funding acquisition, P.S.K. and A.K.-P. All authors contributed equally to this work. All authors have read and agreed to the published version of the manuscript.}

\funding{This research was funded by the Hellenic Foundation for Research and Innovation (HFRI) under the 3rd Call for HFRI PhD Fellowships with grant number 5657 and by the State Scholarships Foundation (IKY) under Act number MIS 5113934.}

\institutionalreview{Not applicable}

\informedconsent{Not applicable}

\dataavailability{Not applicable}

\acknowledgments{The author (P.S.K.) acknowledges support by the State Scholarships Foundation (IKY) under Act number MIS 5113934. The research work was supported by the Hellenic Foundation for Research and Innovation (HFRI) under the 3rd Call for HFRI PhD Fellowships (Fellowship Number: 5657). The authors would like to thank L. Rezzolla for their useful comments and insight.}

\conflictsofinterest{The authors declare no conflict of interest.} 


\abbreviations{Abbreviations}{
The following abbreviations are used in this manuscript:\\

\noindent 
\begin{tabular}{@{}ll}
EOS & Equation of state \\
SN & Supernova \\
GW & Gravitational waves\\
BNS & Binary neutron stars \\
GR &  General relativity\\
TE & Tolman--Ehrenfest\\
TOV & Tolman--Openheimer--Volkoff\\
MDI & Momentum-dependent interaction \\
LHS & Left-hand side \\
RHS & Right-hand side
\end{tabular}}

\begin{adjustwidth}{-\extralength}{0cm}

\reftitle{References}




\end{adjustwidth}
\end{document}